# Microvasculature Segmentation and Inter-capillary Area Quantification of the Deep Vascular Complex using Transfer Learning

Julian Lo[1], Morgan Heisler[1], Vinicius Vanzan[2], Sonja Karst[2,3], Ivana Zadro Matovinovic[4], Sven Loncaric[4], Eduardo V. Navajas[2], Mirza Faisal Beg[1], Marinko V. Sarunic[1]

[1] School of Engineering Science, Simon Fraser University, Canada
[2] Department of Ophthalmology and Visual Sciences, University of British Columbia, Canada
[3] Department of Ophthalmology and Optometry, Medical University of Vienna, Austria
[4] Faculty of Electrical Engineering and Computing, University of Zagreb, Croatia

# Abstract

**Purpose:**
Optical Coherence Tomography Angiography (OCT-A) permits visualization of the changes to the retinal circulation due to diabetic retinopathy (DR), a microvascular complication of diabetes. We demonstrate accurate segmentation of the vascular morphology for the superficial capillary plexus and deep vascular complex (SCP and DVC) using a convolutional neural network (CNN) for quantitative analysis.

**Methods:**
Retinal OCT-A with a 6x6mm field of view (FOV) were acquired using a Zeiss PlexElite. Multiple-volume acquisition and averaging enhanced the vessel network contrast used for training the CNN. We used transfer learning from a CNN trained on 76 images from smaller FOVs of the SCP acquired using different OCT systems. Quantitative analysis of perfusion was performed on the automated vessel segmentations in representative patients with DR.

**Results:**
The automated segmentations of the OCT-A images maintained the hierarchical branching and lobular morphologies of the SCP and DVC, respectively. The network segmented the SCP with an accuracy of 0.8599, and a Dice index of 0.8618. For the DVC, the accuracy was 0.7986, and the Dice index was 0.8139. The inter-rater comparisons for the SCP had an accuracy and Dice index of 0.8300 and 0.6700, respectively, and 0.6874 and 0.7416 for the DVC.

**Conclusions:**
Transfer learning reduces the amount of manually-annotated images required, while producing high quality automatic segmentations of the SCP and DVC. Using high quality training data preserves the characteristic appearance of the capillary networks in each layer.

**Translational Relevance:**
Accurate retinal microvasculature segmentation with the CNN results in improved perfusion analysis in diabetic retinopathy.

# 1 Introduction

Diabetic retinopathy (DR) is a complication of diabetes mellitus, the most common cause of vision loss among people with diabetes, which affects 749,800 Canadians.[1] DR damages the structure of the capillaries in the retina,[2] leading to widespread areas of retinal ischemia as it progresses. Optical coherence tomography angiography (OCT-A) is a rapidly emerging imaging technology that allows for the retinal microvasculature to be seen volumetrically in micrometer-scale detail.[3,4] OCT-A has shown to produce images that closely relate to histology,[5–8] and presents a noninvasive and dye-free alternative with a lower risk of complications[9] when compared to the current gold standard, fluoroscein angiography (FA).

Analysis and quantification of the retinal microvasculature benefits from multi-scale imaging with fields-of-view (FOV's) ranging from ~2x2mm to ~6x6mm. At a smaller FOV, the capillaries that comprise the structure of the microvasculature can be individually resolved, whereas with a larger FOV, macroscopic features, including regions of capillary non-perfusion, can be identified. With recent research hypothesizing that early manifestations of DR form in the retinal periphery,[10] improving the vessel segmentations and tools for quantification in wider fields of view for both the superficial capillary plexus (SCP) and deep vascular complex (DVC)[11] are important assets to clinicians. The DVC resulted from combining the intermediate and deep capillary plexuses due to difficulty in resolving each plexus individually, and has shown a higher correlation to retinal ischemia in DR.[12,13]

OCT-A images are information-rich, and time-consuming for a clinician to trace the vessels for detailed analysis. For the cases of highest clinical interest, small changes in the capillaries need to be detected. Consequently, accurate automated methods of microvasculature segmentation are an essential step towards quantification. However, the efficacy of traditional image processing algorithms can vary based on artifacts present in the image, most notably from noise. Segmentation of microvasculature in funduscopic photos, as well as FA have been examined,[14] however fewer specific algorithms developed for OCT-A have been developed. Simple thresholding of the OCT-A image intensity has been applied,[15,16] but these approaches pose numerous drawbacks in its invariance to microvasculature features and performance when applied to lower-quality images with a low signal-to-noise ratio (SNR). The disadvantages of thresholding can be seen in Figure 1, where an image of the DVC in a patient with mild diabetic retinopathy was processed with Otsu's method.[17] This representative example of thresholding demonstrates that vessels do not maintain continuity, and a significant portion of the speckle is erroneously delineated as a vessel.

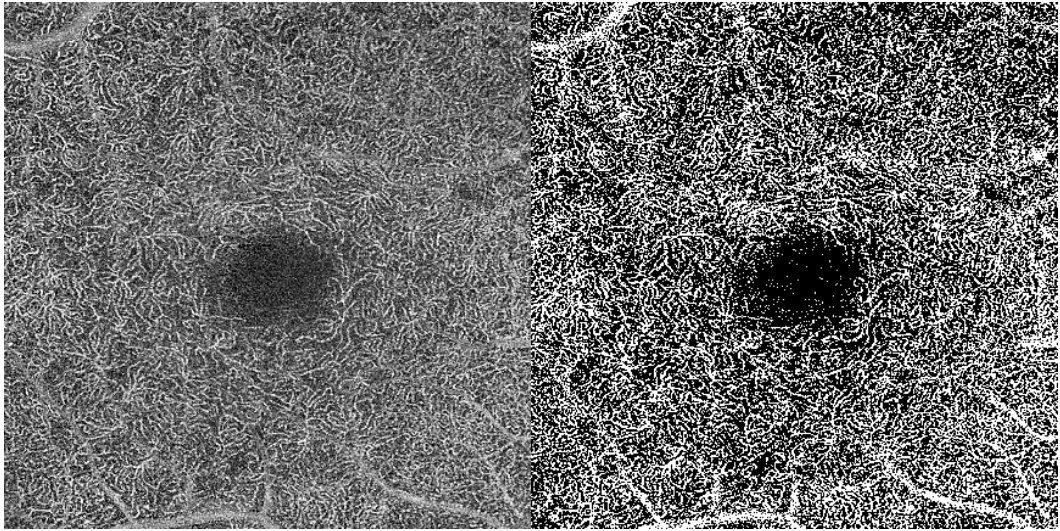

*Figure 1: Left: original single-frame image of the deep vascular complex. Right: image thresholded using Otsu's Method.*

A method using a vesselness filter[18] has been developed, but is similarly limited by SNR. A tophat filter and optimally oriented flux method for segmenting the vessels[19] has been implemented and demonstrated for brain imaging in mice. In addition, some commercial OCT systems also provide segmentation of the vessels but generally face the same issues with images with low SNR.

Machine learning is a rapidly growing field, showing promising results for numerous ophthalmological applications. A few recent reports that are related to this topic include: retinal layer[20–22] and capillary plexus[23] segmentation, cone photoreceptor identification,[24,25] macular fluid segmentation,[26] geographic atrophy segmentation,[27] OCT image categorization,[28] diagnosis and referral for retinal disease patients,[29,30] and synthesis of funduscopic images.[31] Additionally, recent reports have published online tools to improve the accessibility of machine learning-based retinal layer segmentation through intuitive user interfaces that can be used directly by clinicians.[32] Machine learning algorithms have also been applied towards OCT-A segmentation, with a recent approach (MEDnet)[33] applying a convolutional neural network (CNN) to identify and segment avascular areas in wide-field images of the SCP. We have also previously published a method of using a CNN to segment 1x1mm images of the SCP.[34] Machine learning algorithms using CNN's are well suited to address the issues of vessel segmentation through a series of trainable filters. These filters allow the segmentation to be sensitive to vessel boundaries, and hence also have the potential to preserve the vessel widths. However, even with the strengths of machine learning, the quality of the OCT-A images will have a significant impact on the results of the vessel segmentation and quantification.

In our previous work, we proposed a method to register and average multiple sequentially acquired OCT-A images in order to significantly improve image quality and vessel discernibility.[35] Related works in the Literature have also investigated averaging of OCT-A images to improve vessel contrast[36] and automated biomarker identification algorithms.[37] However, this requires prolonged imaging sessions, which is not always possible, particularly in a high-volume clinical environment. Therefore, there is greater clinical utility in the development of an algorithm that can accurately segment, and subsequently quantify, the vasculature and corresponding inter-capillary areas (ICA's) from one single-frame OCT-A image. Quantification of individual ICA's has been previously explored,[15,18] but used 3x3mm images of the SCP.

With many approaches presenting accurate analysis of the SCP, the contribution of this report is to describe an original and novel method to accurately and automatically segment and quantify the DVC. This used an approach of transfer learning, referred to as fine-tuning, for the segmentation of retinal microvasculature in single-frame, wide-field 6x6mm OCT-A images for the purposes of ICA quantification. The developed framework allows for the adaptation of an initial segmentation network to a new dataset with significantly fewer manually graded training examples. We combined the approach of OCT-A averaging to generate high contrast images of the vascular networks with supervised learning to provide the CNN with accurate ground truth data in order to guide the vessel segmentation even in the case of a single (unaveraged) OCT-A image. The computer-generated segmentations were qualitatively examined by retinal specialists and compared to manual segmentations from a trained rater. The outputs of the automated vessel analysis provide near-immediately available, quantitative information on the microvasculature and ICA's from a single OCT-A volume, and hence can potentially accelerate treatment plans and improve DR prognosis.

# 2 Methods

## 2.1 Subject Criteria and Data Preparation

Subject recruitment and imaging took place at the Eye Care Centre of Vancouver General Hospital, and North Shore Eye Associates. The project protocol was approved by the Research Ethics Boards at the University of British Columbia and Vancouver General Hospital, and the experiment was performed in accordance with the tenets of the Declaration of Helsinki. Written informed consent was obtained by all subjects.

Subjects in the control group (n = 8) displayed no evidence of retinal or ocular pathology upon examination by an experienced retina specialist. Subjects classified as diabetic (n = 28) were diagnosed with DR based on the international DR severity scale.[38] All subjects were screened for clear ocular media, ability to fixate, and ability to provide informed consent before imaging. In addition, patients with diabetic macular edema were not included in the study.

Table 1: Demographics of the control dataset used in this study.

| Gender | n | Mean Age (Standard Deviation) |
|---|---|---|
| Male | 4 | 24.5 (3) |
| Female | 4 | 53.5 (23.1) |

## 2.2 Optical Coherence Tomography Instrumentation

The data used for this study was acquired with the ZEISS PlexElite (Carl Zeiss Meditec, Dublin, CA) with software version 1.7.31492. The nominal 6x6mm scanning protocol was used, sampling at a 512x512 resolution at a rate of 100,000 A-scans per second at a visual angle of 20.94 degrees. Each B-scan was repeated 4 times at the same position and the optical micro-angiography (OMAG) implemented on the commercial imaging system was used to generate the angiographic information. The A-scan depth was 3mm with an axial resolution of 6.3 μm and a transverse resolution of 20 μm, as described in the product specifications.

The inner limiting membrane (ILM) and posterior boundary of the outer plexiform layer (OPL) were used as the segmentation boundaries for the commercial device. The SCP and DVC were extracted, with projection artifacts removed, via a built-in software feature in the Zeiss PlexElite. Scans were only included in the study if the system specified signal strength was 8 (out of 10) or higher.

## 2.3 Network Architecture

The network for vessel segmentation used a variation of the U-Net[39] architecture, which was adapted for two classes: vessel and background. The basic U-Net architecture is shown in Figure 2 and consists of convolutional and pooling layers. The convolutional layers consist of a series of trainable filters, which are correlated across the image and subsequently passed through a rectifier linear unit activation with units capped at 6 (ReLU-6).[40] Each convolutional layer was followed by a batch normalization layer, as well as a dropout layer with a coefficient of 0.5.[40] Pooling layers were inserted to increase the receptive field of the subsequent filters in the convolutional layers, helping with generalization to prevent overfitting.

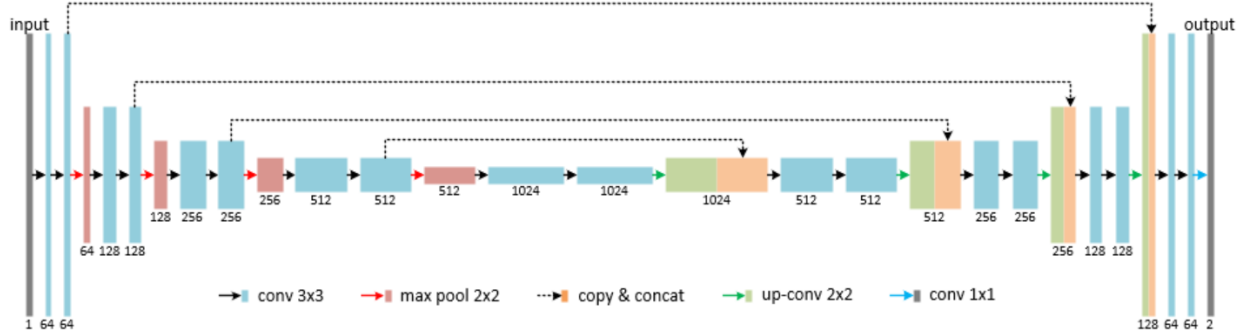

*Figure 2: U-net architecture.*

The U-Net architecture provides high-resolution feature extraction through its structure, which consists of a contracting and expanding path. For each level in the contracting path, high resolution weights are combined with the upsampled, generalized weights (span connections) in the corresponding level of the expanding path. This allows the network to retain the learned localization information and better segment smaller, more detailed structures in the image. In addition, methods using residual blocks for each convolutional block were experimented, however this resulted in similar or lower performance (data not shown).

## 2.4  Training

Two OCT-A datasets were used for training the network. To construct the initial weights, data were acquired from a previous study.[41] Briefly, this data consisted of 29 images with a 2x2mm FOV acquired with a prototype swept source OCT instrument[42] and 47 images with a 3x3mm FOV acquired with a commercial spectral domain OCT instrument. Each OCT-A image was manually segmented using a Microsoft Surface Pro tablet and GNU image manipulation program (GIMP) by one trained rater and verified and accepted by two additional trained raters.

To construct the initial weights, the network hyperparameters were optimized using 3-fold cross-validation. This resulted in a network trained over 120 epochs using the Adam optimizer, with an initial learning rate of $10^{-4}$, and a custom epsilon value of $10^{-5}$. Evaluation was performed qualitatively on a set of acquired 3x3mm FOV OCT-A images across all devices based on images most recently acquired at the clinic. Segmentation of a single 3x3mm or 2x2mm image using the network took approximately two seconds on a Nvidia RTX 2060 GPU, with a possible decrease to 0.3 seconds per image when segmented in larger batches of ~10 images.

The network with the initial weights was subsequently fine-tuned on the 6x6mm FOV images acquired with the PlexElite. This was done in two stages: First, a smaller dataset of 10 single-frame OCT-A images of each of the SCP and DVC for which there existed a corresponding high-quality averaged image was identified. As described in our previously-published study,[35] images were registered and averaged based on a template image that was free of microsaccadic motion. This allowed us to use the averaged OCT-A images to construct ground-truth labels for each single-frame template image. These labeled ground truth vessel segmentations were subsequently paired with the single-frame template OCT-A images to train the deep neural network to perform segmentations approaching the quality of averaged images, while only using single-frame images.

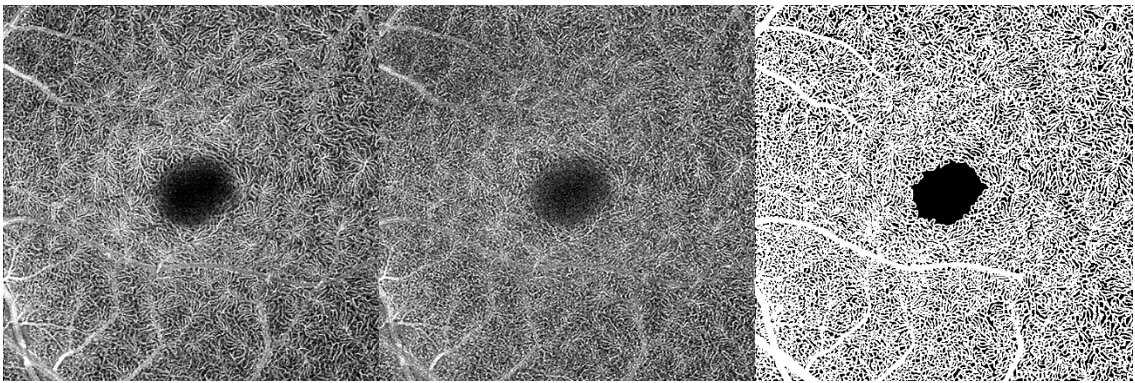

*Figure 3: Left: 10-frame averaged 6x6mm image of the deep vascular complex (DVC). Centre: Single-frame template image of the same region. Right: Manually-segmented averaged image, to be paired with the template image for training.*

For the SCP, the automated segmentations of the averaged images generated by the network with initial weights were adequate, as determined by a separate group of trained raters; hence these results were fed back into the network as additional training examples. However, the automated segmentations on the DVC required additional manual correction due to the lower SNR of the images in this layer and the different morphological features of the vasculature relative to the SCP. Using the initial weights as a guideline for the manual raters would introduce biases that could negatively impact further stages of training; hence, the vasculature of the DVC images was instead first segmented through Otsu's method.[17] Another masked and trained rater manually corrected the resulting segmentation using a Microsoft Surface Pro tablet and GIMP. All segmentations were reviewed and accepted by two of three other trained raters. A second trained rater segmented three images to obtain inter-rater metrics.

Due to memory limitations when training, each 6x6 image was separated into 4 quadrants, which were saved as 4 separate images. The same method of augmentation and cross-validation was used. This resulted in the network being trained through the Adam optimizer, using an initial learning rate of $10^{-2}$, and a custom epsilon value of $10^{-2}$. The forward inference segmentation of a single 6x6mm image using the network took approximately four seconds, with a possible decrease to 0.5 seconds per image when segmented in larger batches.

To further reinforce the manually-segmented dataset, extensive data augmentation was performed. Each OCT-A image (along with its corresponding manual segmentation) in the training set was rotated 90 degrees three times with no processing. Next, to account for noise, each image was rotated 90 degrees an additional 5 times with various contrast adjustments, which included contrast-limited adaptive histogram equalization, as well as the built-in imadjust function in MATLAB. To account for motion, each rotated image was also separated into randomly-sized strips, which were re-ordered randomly to simulate motion artifacts in the image. The probability maps resulting from the automated segmentations were binarized at a value of 0.5, the default class cut-off in the probability map. After binarization, isolated clusters of less than 30 pixels were deemed false positives and removed.

Additional training data were required to improve the performance of the network on the DVC using a single frame OCT-A image. The intermediate network (after training on the 10 manually-segmented images) was able to segment additional 6x6mm averaged OCT-A images of the DVC. Subsequently, the next stage of fine-tuning the CNN involved using this intermediate network to generate additional training data, using manual inspection for quality, but not requiring laborious manual segmentation at the scale of capillaries. We applied the intermediate network trained on the manual segmentations to all 50 averaged OCT-A images in our 6x6mm dataset. After manual inspection, we identified 39 images for each of the SCP and DVC with adequate automated segmentations that constituted the new training set. A summary of the training sets used is provided in Table 2.

Similarly to the first stage, the weights were initialized with the original network trained with the first dataset of 2x2mm and 3x3mm images. The same methods of image augmentation and training were applied, and cross-validation resulted in an initial learning rate of $10^{-2}$, and a custom epsilon value of $10^{-2}$ using the Adam optimizer.

*Table 2: Overview of the three datasets used to train the fine-tuned network*

|  | **2x2/3x3mm dataset** | **First 6x6mm dataset** | **Second 6x6mm dataset** |
|---|---|---|---|
| **Training images** | SCP: 76 averaged and single-frame | SCP: 10 single-frame DVC: 10 single-frame | SCP: 39 single-frame DVC: 39 single-frame |
| **Ground-truth segmentations** | SCP: manual | SCP: automated DVC: manual | SCP: automated DVC: automated |

## 2.5 Performance Evaluation

To evaluate the automated segmentation performance, a number of metrics were calculated. The number of true positives (TP), false positives (FP), false negatives (FN), and true negatives (TN) were calculated using pixel-wise comparison between a reference manual segmentation and the thresholded binary output of the neural network. To calculate inter-rater metrics, these metrics were calculated by comparing one manual segmentation to another by a different rater. In the context of this study, pixels corresponding to vessels and the background comprised the positive and negative classes, respectively. Using the TP, FP, FN, and TN numbers we can calculate the accuracy of the segmentation, as shown in Equation 1:

$$Accuracy = \frac{TP + TN}{TP + TN + FP + FN}. \tag{1}$$

Additionally, we can compute the Dice similarity coefficient (DSC), which quantifies the similarity between two segmented images, through measuring the degree of spatial overlap. The DSC value ranges from 0, indicating no spatial overlap, and 1, indicating complete overlap and is calculated by Equation 2:

$$DSC = \frac{2TP}{2TP + FP + FN}. \tag{2}$$

Three methods were tested: using only the initial weights, using a network solely trained on the 6x6mm dataset, and the fine-tuned network, trained on both datasets.

## 2.6 Post-Processing of the Automated Vessel Segmentation

The neural network generated segmentation of the vessels in the OCT-A images, but further processing was required for quantitative analysis. Next, the ICA's were identified, determined by the non-vessel pixels. The largest ICA within a small region in the center of the image was defined as the foveal avascular zone (FAZ). All erroneously segmented pixels within the FAZ were set to a non-vessel classification, with the centroid then used as the center of the Early Treatment of Diabetic Retinopathy Study (ETDRS) grid.

Two metrics were of interest when quantifying ICA's: the area of the region and maximum ischemic point (MIP), defined as the point of maximum distance to nearest vessel within the ICA. The metrics were measured for each ETDRS region and compared to a database of healthy eyes, for which the SCP and DVC were extracted. As outlined in Section 2.1, 8 healthy controls were recruited, resulting in a possible 16 eyes. Of these, we were able to obtain high-quality averaged images for 12 to construct the reference database and were also included in the training dataset outlined in Table 2. Each measured ICA was color-coded and overlaid on the original image based on number of standard deviations from the mean. Perifoveal vessel density (for each ETDRS region) was also calculated as the proportion of measured area occupied by pixels which were classified by the algorithm as a vessel. In addition, the projection artifacts remaining in segmentations of the DVC were excluded for the calculation of vascular metrics including vessel density and vessel index. This was done through an automated MATLAB post-processing step using image erosion and dilation.

# 3 Results

## 3.1 Network Performance Evaluation

### 3.1.1 Quantitative Segmentation Comparison

Tables 3 and 4 show comparative quantitative results when segmenting the SCP and DVC respectively. The network trained on the initial dataset of 2x2mm and 3x3mm images, the network solely trained on the 6x6mm dataset, and the network trained with our proposed transfer-learning method were enumerated as Network A, B, and C, respectively. The accuracy and Dice index for Network C showed a high similarity between segmentations of the single-frame template images and the averaged images. The inter-rater metrics were only conducted on the manually-segmented datasets and are intended to be a representative number illustrating the difficulty of this problem and the variation in the metrics between human raters. Table 5 shows the same networks but evaluated on the original 2x2mm and 3x3mm dataset.

*Table 3: Comparative quantitative results of the segmentation of the SCP between three networks: Network A consisted of only the initial weights, Network B was trained on only the images from the 6x6mm dataset, and Network C was the fine-tuned network using our proposed transfer learning method.*

|  | Network A | Network B | Network C | Inter-rater |
|---|---|---|---|---|
| **Accuracy** | 0.8141 | 0.8534 | 0.8599 | 0.8300 |
| **Dice similarity index** | 0.8060 | 0.8586 | 0.8618 | 0.6700 |

*Table 4: Comparative quantitative results of the segmentation of the DVC between three networks: Network A consisted of only the initial weights, Network B was trained on only the images from the 6x6mm dataset, and Network C was the fine-tuned network using our proposed transfer learning method.*

|  | Network A | Network B | Network C | Inter-rater |
|---|---|---|---|---|
| **Accuracy** | 0.6934 | 0.7822 | 0.7986 | 0.6874 |
| **Dice similarity index** | 0.6469 | 0.8065 | 0.8139 | 0.7416 |

*Table 5: Comparative quantitative results of the segmentation of the 2x2mm and 3x3mm dataset between three networks: Network A consisted of only the initial weights, Network B was trained on only the images from the 6x6mm dataset, and Network C was the fine-tuned network using our proposed transfer learning method.*

|  | Network A | Network B | Network C | Inter-rater |
|---|---|---|---|---|
| **Accuracy** | 0.8677 | 0.8329 | 0.8350 | 0.8300 |
| **Dice similarity index** | 0.8395 | 0.8059 | 0.8066 | 0.6700 |

### 3.1.2 Qualitative Segmentation Comparison

The fine-tuned network was qualitatively evaluated on data unseen by the CNN during training on control and DR patients. Figure 4 focuses on a peripheral area of the SCP located close to the optic nerve head, where the elongated vascular structure of retinal peripapillary capillaries (RPC's) are visible. As shown in Figure 4-C2, Network C (the fine-tuned network) preserves the features characteristic of the RPC's, and segments larger vessels more accurately than Network A (the initial weights). The differences between Network B (trained solely with the 6x6mm dataset) and Network C are less pronounced, due to the higher SNR present in images of the SCP.

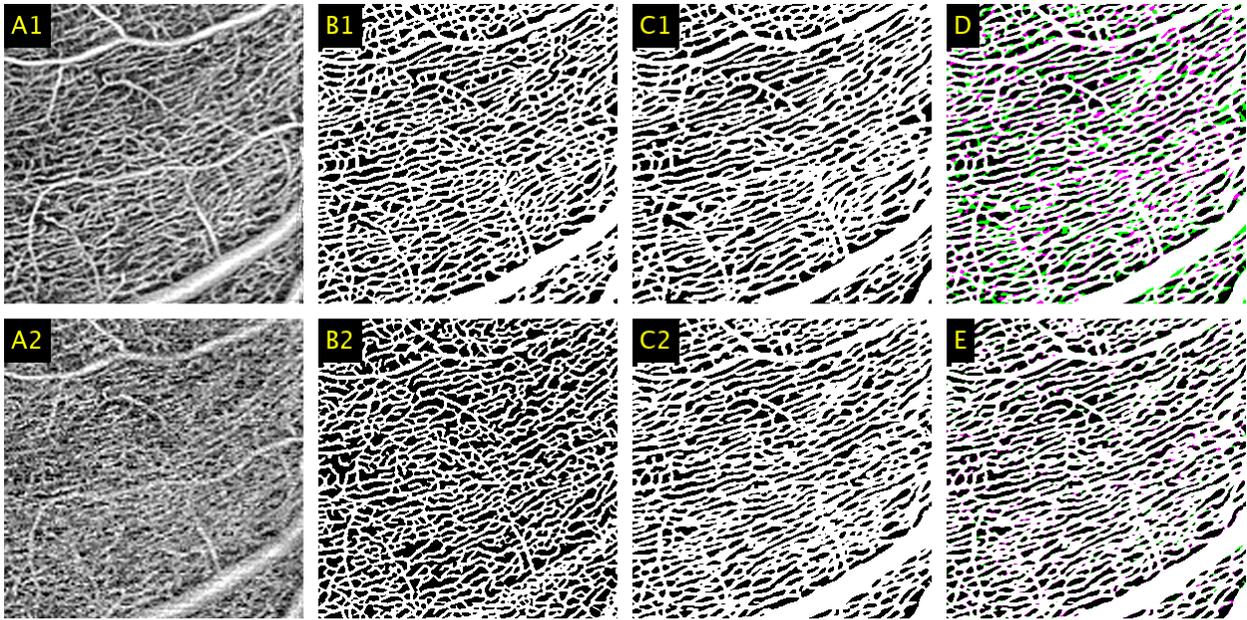

*Figure 4: A1: 2x2mm window of an averaged 6x6mm image taken of the superficial capillary plexus (SCP). A2: 2x2mm window of the corresponding region in the equivalent single-frame template image. B1: averaged image segmented using the initial weights (Network A). B2: single-frame image segmented using Network A. C1: averaged image segmented using the fine-tuned network (Network C). C2: single-frame image segmented using Network C. D: comparison between the automated segmentations between the averaged and single-frame images produced by Network C, represented by magenta and green respectively, with white representing the union. E: comparison between single-frame segmentations between Network B and C, represented by magenta and green respectively, with white representing the union.*

Figure 5 shows an additional enlarged comparison for an image of the SCP. It can be seen here that Network C is able to accurately segment the areas of ischemia observed in the averaged images, and as also shown in Figure 4, segments larger vessels more accurately.

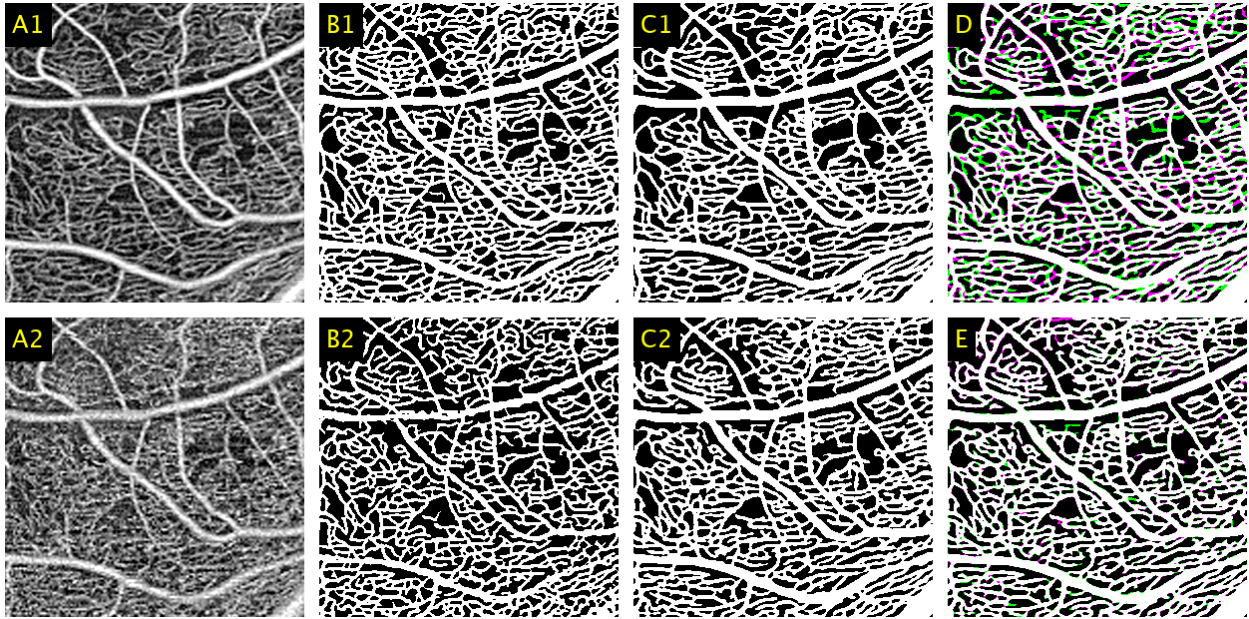

*Figure 5: A1: 2x2mm window of an averaged 6x6mm image taken of the superficial capillary plexus (SCP). A2: 2x2mm window of the corresponding region in the equivalent single-frame template image. B1: averaged image segmented using the initial weights (Network A). B2: single-frame image segmented using Network A. C1: averaged image segmented using the fine-tuned network (Network C). C2: single-frame image segmented using Network C. D: comparison between the automated segmentations between the averaged and single-frame images produced by Network C, represented by magenta and green respectively, with white representing the union. E: comparison between single-frame segmentations between Network B and C, represented by magenta and green respectively, with white representing the union.*

Figure 6 shows an enlarged comparison of the segmentations results obtained by Network A and Network C when segmenting the elongated, lobular capillary structure of the DVC in a lower-quality image. As shown in Figure 6-B2, certain clusters of vessels were erroneously treated as noise by Network A, resulting in regions of false negatives. This is characteristic of single-frame OCT-A images; the blurred-out regions were replaced by a discernible vessel structure when using the averaged OCT-A images. The results presented in Figure 6-C2 are representative of the outputs from Network C, which eliminated a portion of these false negatives.

As mentioned in Section 2.1, it is important to note that the thicker vessels shown traversing horizontally in the samples in Figure 6 do not exist in the DVC and are residual projection artifacts from the large arteries of the SCP that obscure the capillaries underneath. These are subsequently automatically removed as a post-processing step when calculating the vessel density and vessel index metrics. When segmenting with Network A, these projection artifacts were erroneously segmented as additional capillaries. This can also be seen in Figure 6-E where the green spots along the thicker vessels indicate that the segmentation produced by Network B is not as continuous as Network C. With accurate and continuous projection artifact delineation, these can be more accurately removed in post-processing.

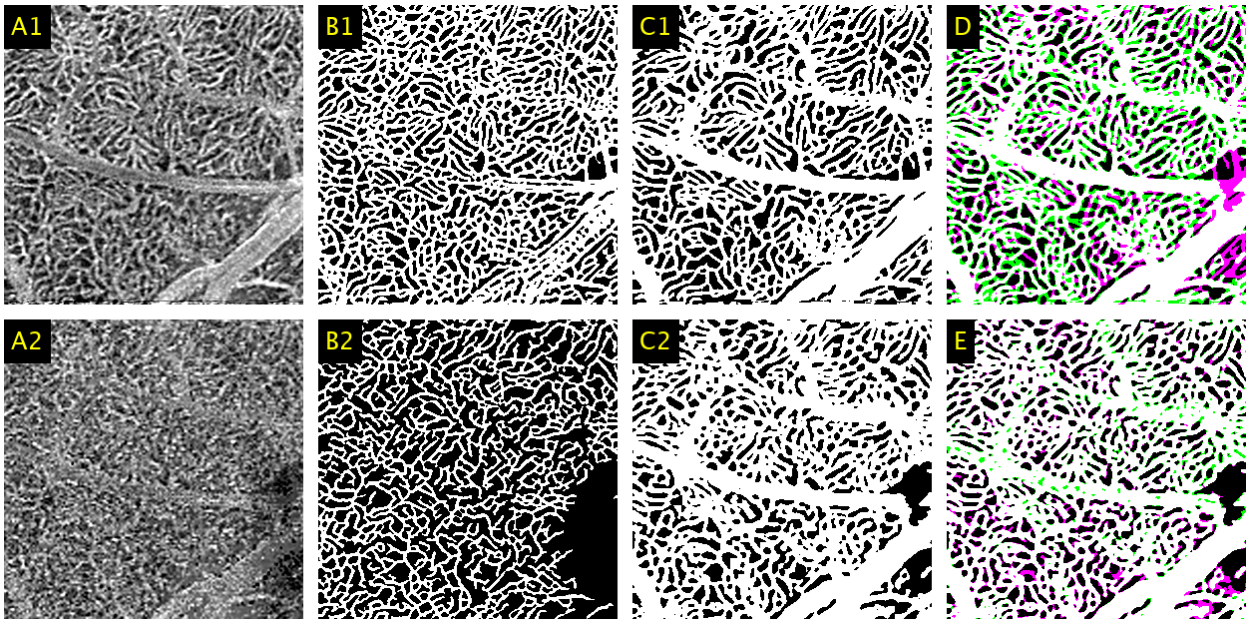

*Figure 6: A1: 2x2mm window of an averaged 6x6mm image taken of the deep vascular complex (DVC). A2: 2x2mm window of the corresponding region in the equivalent single-frame template image. B1: averaged image segmented using the initial weights (Network A). B2: single-frame image segmented using Network A. C1: averaged image segmented using the fine-tuned network (Network C). C2: single-frame image segmented using Network C. D: comparison between the automated segmentations between the averaged and single-frame images produced by Network C, represented by magenta and green respectively, with white representing the union. E: comparison between single-frame segmentations between Network B and C, represented by magenta and green respectively, with white representing the union.*

Figure 7 shows an additional enlarged comparison of the results when segmenting the DVC with different versions of the deep neural network. The images segmented by Network C, shown in Figure 7-C1 and Figure 7-C2, more closely follow the elongated, lobular ICA morphology of the DVC and the results are less prone to over-segmenting noise. This presents a substantial improvement over the images segmented by Network A, as shown in Figure 7-B1 and Figure 7-B2, the results of which incorrectly apply the branching structure characteristic of the SCP to the DVC.

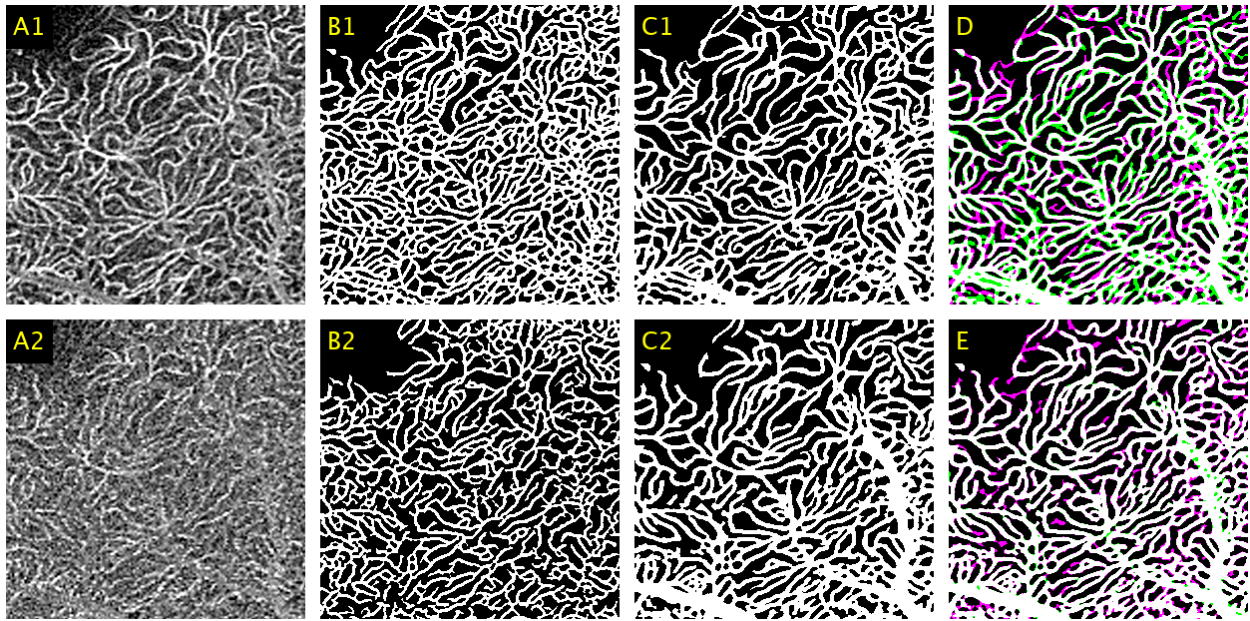

*Figure 7: A1: 2x2mm window of an averaged 6x6mm image taken of the deep vascular complex (DVC). A2: 2x2mm window of the corresponding region in the equivalent single-frame template image. B1: averaged image segmented using the initial weights (Network A). B2: single-frame image segmented using Network A. C1: averaged image segmented using the fine-tuned network (Network C). C2: single-frame image segmented using Network C. D: comparison between the automated segmentations between the averaged and single-frame images produced by Network C, represented by magenta and green respectively, with white representing the union. E: comparison between single-frame segmentations between Network B and C, represented by magenta and green respectively, with white representing the union.*

## 3.2 Inter-capillary Area Evaluation

Figure 8 shows representative images, segmentations, and standard deviation maps for diabetic subjects without DR, with mild/moderate non-proliferative NPDR, and with severe non-proliferative DR as graded by a retina specialist.

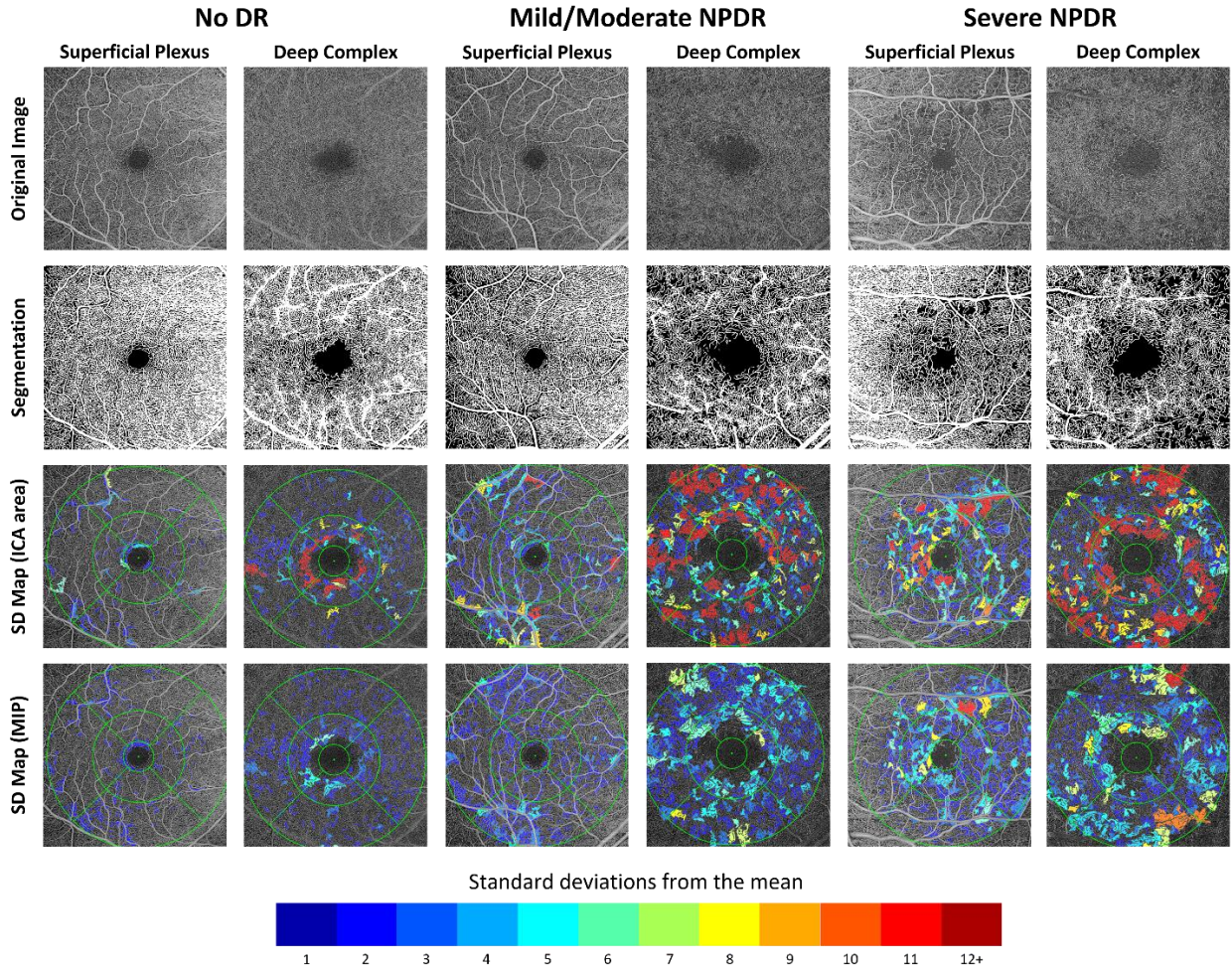

*Figure 8: Labeled standard deviation maps for subjects with no diabetic retinopathy, mild/moderate non-proliferative diabetic retinopathy, and severe non-proliferative diabetic retinopathy. Original images of the deep vascular complex have been brightened for clarity. All inter-capillary areas are labeled based on the number of standard deviations its maximum ischemic point and area exceeded the reference mean.*

# 4 Discussion

Early detection of DR is paramount to ensuring effective treatment and improved patient quality-of-life. Detecting changes in both the SCP and DVC have been identified as potential early biomarkers of DR. As a result, accurate segmentation and quantification of increasingly wide-field images of both the SCP and DVC will allow for further insight into the emergence and progression of DR.

We designed a transfer learning-based framework for automated segmentation of the microvasculature in the SCP and DVC, as well as quantification of the ICA's in 6x6mm single-frame OCT-A images. The framework consists of two convolutional neural networks: an initial network, trained on 2x2mm and 3x3mm images; and a second network, which utilized the pre-existing weights and fine-tuning on a smaller dataset of 6x6mm images of both the SCP and DVC. This approach allowed for accurate feature detection despite a limited training set, with results that exceed the intra-rater accuracy. In particular, fine-tuning from an existing set of data provided more robust projection artifact delineation in the DVC, allowing for removal in post-processing when computing vascular metrics. The resulting ICA quantifications allow for a closer investigation into suspected areas of low perfusion but does not expressly define what constitutes such areas.

A prevailing limitation of many machine learning problems is training dataset acquisition. For our study, manually segmenting an individual 6x6mm image of the DVC took each rater an average of 4 hours to complete, which can pose a significant challenge for problems requiring larger datasets. Solely training a new network on our limited, manually-segmented, 6x6mm dataset would overfit to the training set and including this new dataset with the original would result in a heavy data imbalance. The introduction of additional automated segmentations of averaged images greatly increased the size and quality of our training set, from 10 images of each of the SCP and DVC to 39. This allowed for a larger variation in training samples, consequently improving network performance.

The impact of the training examples is most evident in the segmentations of the DVC, where the initial weights produced segmentations that significantly differed from the images produced by the fine-tuned network. As seen in Figure 6 and Figure 7, vessels segmented by the initial weights closely resemble the denser morphological characteristics of the SCP. In particular, the ICA's in the DVC follow a lobular pattern, which is reflected more accurately in the segmentations generated by the fine-tuned network.

Another limitation is the quality of the data. Currently, images with an SSI of 8 or lower, as well as images with excessive microsaccadic eye motion were omitted from the study. If there are excessive microsaccadic artifacts, microvascular features begin to blur, and can be subsequently classified as noise by the network. This emphasizes the importance of using the averaged 6x6mm images as the ground-truth data obtained from manual segmentations because it will be the most anatomically accurate. Our previously-published method of averaging and registering single-frame images based on a template[35] allowed for segmentations of averaged images to be paired with single-frame training data, greatly improving the quality of our training samples. Segmentation quality appeared to be independent of location in the image, as automated segmentation accuracy was consistent across the 6x6mm FOV in the absence of additional artifacts.

To summarize, we designed a machine learning framework to accurately segment and quantify the retinal microvasculature in the SVP and the DVC. It produces immediately-available segmentations, which provide clinicians with a tool for in-depth analysis of ICA's and the level of retinal perfusion. Through this framework, patient care for DR can be adapted individually, improving compliance and patient prognosis. In addition, visualization and quantification of retinal vasculature at a high level of accuracy provide more information about disease activity and therefore may add to individualized patient care.

# 5 List of Abbreviations

| | |
|---|---|
| DR | Diabetic retinopathy |
| OCT-A | Optical coherence tomography angiography |
| FA | Fluoroscein angiography |
| FOV | Field-of-view |
| SCP | Superficial capillary plexus |
| DVC | Deep vascular complex |
| SNR | Signal-to-noise ratio |
| CNN | Convolutional neural network |
| ICA | Inter-capillary area |
| OMAG | Optical micro-angiography |
| ILM | Inner limiting membrane |
| OPL | Outer plexiform layer |
| GIMP | GNU Image Manipulation Program |
| TP | True positive |
| FP | False positive |
| FN | False negative |
| TN | True negative |
| FAZ | Foveal avascular zone |
| ETDRS | Early Treatment of Diabetic Retinopathy Study |
| MIP | Maximum ischemic point |
| RPC | Retinal peripapillary capillary |

# 6 Acknowledgements

**Funding:** This work was supported by funding from the following agencies: Brain Canada, Diabetes Action Canada, National Sciences and Engineering Research Council of Canada, Canadian Institutes of Health Research, Alzheimer Society Canada, Michael Smith Foundation for Health Research, and Genome British Columbia.

**Disclosures:** J. Lo: None; M. Heisler: None; V. Vanzan: None; S. Karst: None; I. Z. Matovinovic: None; S. Loncaric: None; E. V. Navajas: None; M. F. Beg: None; M. V. Sarunic: Seymour Vision, Inc. (I)